\documentclass[a4paper]{article}

\usepackage{INTERSPEECH2020}

\usepackage{booktabs} %
\usepackage{multirow} %
\usepackage{diagbox} %
\usepackage{color,url,cite}
\usepackage{enumerate}
\usepackage[ruled,linesnumbered]{algorithm2e}
\usepackage[labelfont=bf]{caption} %

\usepackage[acronym]{glossaries}
\newacronym{asr}{ASR}{automatic speech recognition}
\newacronym{blstm}{BLSTM}{bidirectional long-short term memory}
\newacronym{blstmp}{BLSTMP}{bidirectional long-short term memory with projection}
\newacronym{casa}{CASA}{computational auditory scene analysis}
\newacronym{cer}{CER}{character error rate}
\newacronym{cirm}{cIRM}{complex ideal ratio mask}
\newacronym{cnn}{CNN}{convolutional neural network}
\newacronym{ctc}{CTC}{connectionist temporal classication}
\newacronym{dct}{DCT}{discrete cosine transform}
\newacronym{dft}{DFT}{discrete {Fourier} transform}
\newacronym{dnn}{DNN}{deep neural network}
\newacronym{doa}{DOA}{direction of arrival}
\newacronym{dpcl}{DPCL}{deep clustering}
\newacronym{e2e}{E2E}{end-to-end}
\newacronym{fft}{FFT}{fast {Fourier} transform}
\newacronym{gan}{GAN}{generative adversarial network}
\newacronym{gcc}{GCC}{generalized cross correlation}
\newacronym{gcc-phat}{GCC-PHAT}{generalized cross-correlation phase transform}
\newacronym{gmm}{GMM}{{Gaussian} mixture model}
\newacronym{hmm}{HMM}{hidden {Markov} model}
\newacronym{ibm}{IBM}{ideal binary mask}
\newacronym{irm}{IRM}{ideal ratio mask}
\newacronym{ipd}{IPD}{interaural phase difference}
\newacronym{lstm}{LSTM}{long-short term memory}
\newacronym{mfcc}{MFCC}{{Mel}-frequency cepstral coefficient}
\newacronym{mfccs}{MFCCs}{{Mel}-frequency cepstral coefficients}
\newacronym{mimo}{MIMO}{multi-input multi-output}
\newacronym{mlp}{MLP}{multilayer perceptron}
\newacronym{mse}{MSE}{mean square error}
\newacronym{mvdr}{MVDR}{minimum variance distortionless response}
\newacronym{mpdr}{MPDR}{minimum power distortionless response}
\newacronym{nmf}{NMF}{non-negative matrix factorization}
\newacronym{pca}{PCA}{principal component analysis}
\newacronym{pesq}{PESQ}{perceptual evaluation of speech quality}
\newacronym{pit}{PIT}{permutation invariant training}
\newacronym{plp}{PLP}{perceptual linear prediction}
\newacronym{psd}{PSD}{power spectral density}
\newacronym{relu}{ReLU}{rectified linear unit}
\newacronym{rir}{RIR}{room impulse response}
\newacronym{rirs}{RIRs}{room impulse responses}
\newacronym{rnn}{RNN}{recurrent neural network}
\newacronym{s2s}{S2S}{sequence-to-sequence}
\newacronym{sdr}{SDR}{signal-to-distortion ratio}
\newacronym{snr}{SNR}{signal-to-noise ratio}
\newacronym{srp-phat}{SRP-PHAT}{steered-response power phase transform}
\newacronym{stft}{STFT}{short-time {Fourier} transform}
\newacronym{stoi}{STOI}{short-time objective intelligibility}
\newacronym{svm}{SVM}{support vector machine}
\newacronym{tts}{TTS}{text-to-speech}
\newacronym{vad}{VAD}{voice activity detection}
\newacronym{wer}{WER}{word error rate}
\newacronym{wpe}{WPE}{weighted prediction error}
\newacronym{wpd}{WPD}{weighted power minimization distortionless response} 

\newcommand{\revised}{\textcolor{black}}
\ninept

\usepackage[hidelinks,bookmarks=true,hypertexnames=true]{hyperref}

\usepackage{caption} 
\DeclareMathOperator*{\argmin}{arg\,min}

\newcommand{\formattitle}[1]{\textbf{#1}}

\renewenvironment{thebibliography}[1]{
  \begin{oldthebibliography}{#1}
  \setlength{\itemsep}{0.12em}
  \setlength{\parskip}{0.12em}
}
{
  \end{oldthebibliography}
}

\title{
End-to-End Far-Field Speech Recognition with\\Unified Dereverberation and Beamforming
}
\name{Wangyou Zhang$^{1}$, Aswin Shanmugam Subramanian$^{2}$, Xuankai Chang$^{2}$, \\Shinji Watanabe$^{2}$, Yanmin Qian$^{1}$ \thanks{$^\dag$Shinji Watanabe and Yanmin Qian are the corresponding authors.}
}
\address{
    $^1$MoE Key Lab of Artificial Intelligence \& SpeechLab, Department of Computer Science and Engineering, AI Institute, Shanghai Jiao Tong University, Shanghai\\
    $^2$Center for Language and Speech Processing, Johns Hopkins University, USA
}
\email{wyz-97@sjtu.edu.cn, \{aswin, xchang14, shinjiw\}@jhu.edu, yanminqian@sjtu.edu.cn}

\begin{document}
\setlength{\abovedisplayskip}{2pt}
\setlength{\belowdisplayskip}{2pt}
\setlength{\textfloatsep}{5pt plus 0.0pt minus 2.0pt}
\setlength{\floatsep}{7pt plus 0.0pt minus 2.0pt}
\setlength{\intextsep}{7pt plus 0.0pt minus 2.0pt}

\maketitle
\begin{abstract}
Despite successful applications of end-to-end approaches in multi-channel speech recognition, the performance still degrades severely when the speech is corrupted by reverberation.
In this paper, we integrate the dereverberation module into the end-to-end multi-channel speech recognition system and explore two different frontend architectures.
First, a multi-source mask-based weighted prediction error (WPE) module is incorporated in the frontend for dereverberation.
Second, another novel frontend architecture is proposed, which extends the weighted power minimization distortionless response (WPD) convolutional beamformer to perform simultaneous separation and dereverberation. 
We derive a new formulation from the original WPD, which can handle multi-source input, and replace eigenvalue decomposition with the matrix inverse operation to make the back-propagation algorithm more stable.
The above two architectures are optimized in a fully end-to-end manner, only using the speech recognition criterion.
Experiments on both spatialized wsj1-2mix corpus and REVERB show that our proposed model outperformed the conventional methods in reverberant scenarios.
\end{abstract}

\vspace{-.8em}
\section{Introduction}
\label{sec:intro}
Over the past few years, thanks to the advances in deep learning, significant progress has been made in \gls{asr}. Both \gls{dnn}/\gls{hmm} hybrid systems and \gls{e2e} systems have attained surprisingly good performance in close-talk scenarios \cite{Deep-Hinton2012,xiong2018microsoft,chiu2018state,karita2019comparative}.
However, it is still a challenging task to recognize speech signals in far-field scenarios, where background noise and reverberation are commonly observed and even interfering speech from other speakers is involved \cite{haeb2019speech,watanabe2020chime}.
In recent years, many studies have been focusing on the far-field speech recognition task, including the combination of the speech enhancement frontend and \gls{asr} backend \cite{Single-Isik2016, Multi-Yoshioka2018} and noise robust adaptation approaches \cite{Joint-Narayanan2014, Adaptive-Tan2018}.
Meanwhile, it is commonly observed that speech processing with multiple microphones usually outperforms the single-microphone one, because additional spatial information can be exploited. Therefore, many existing microphone array signal processing methods can be utilized as the frontend for end-to-end far-field speech recognition, such as the multi-channel Weiner filter \cite{Spatially-Spriet2004, Optimal-Souden2009}, \gls{mvdr} and \gls{mpdr} beamforming \cite{Optimum-Van2004, Optimal-Souden2009}, multi-frame beamforming \cite{Sequential-Wang2019}, etc.
In addition, reverberation is also an important problem in real scenarios, which can lead to dramatic degradation in the \gls{asr} performance \cite{Distant-Wolfel2009}. Various deep learning based methods have been proposed for dereverberation, including \gls{dnn} based approaches \cite{Neural-Kinoshita2017, Joint-Heymann2019, Generalized-Taniguchi2019} incorporating the \gls{wpe} algorithm \cite{Speech-Yoshioka2010, Generalization-Yoshioka2012} and complex ideal ratio mask based approach for denoising and dereverberation \cite{Time-Williamson2017}.

In this work, we propose a novel \gls{e2e} architecture that can perform dereverberation, beamforming and recognition simultaneously. Inspired by the recently developed unified convolutional beamformer for simultaneous denoising and dereverberation, named \gls{wpd} \cite{Unified-Nakatani2019, Simultaneous-Nakatani2019}, we reformulate \gls{wpd} by replacing eigenvalue decomposition with an equivalent matrix inverse operation, which makes it differentiable and more stable.
The new architecture consists of a frontend and an \gls{asr} backend. In the frontend, two novel architectures are explored for joint speech dereverberation, enhancement and separation. In the backend, a joint \gls{ctc} / attention-based encoder-decoder model \cite{Joint-Kim2017} is used to recognize each separated speech stream.
Note that our proposed framework can be used for both single-speaker and multi-speaker scenarios. And in this paper, we mainly focus on the multi-speaker case, which is a more difficult task.
It is worth noting that this end-to-end architecture is optimized only based on the final \gls{asr} criterion, which was also proven feasible in previous works \cite{Joint-Heymann2019, MIMO-Chang2019, End-Chang2020, Speech-Subramanian2019}.
Our experiments show that our newly proposed method outperformed the conventional end-to-end ASR systems \cite{MIMO-Chang2019, End-Chang2020,Subramanian-arxiv-2019} in both single-speaker and multi-speaker reverberant conditions.

\vspace{-.7em}
\section{End-to-End Multi-Channel ASR}
\label{sec:MIMO-Speech}

\begin{figure}[t]
  \begin{minipage}[b]{\linewidth}
    \centering
    \centerline{\includegraphics[width=\textwidth]{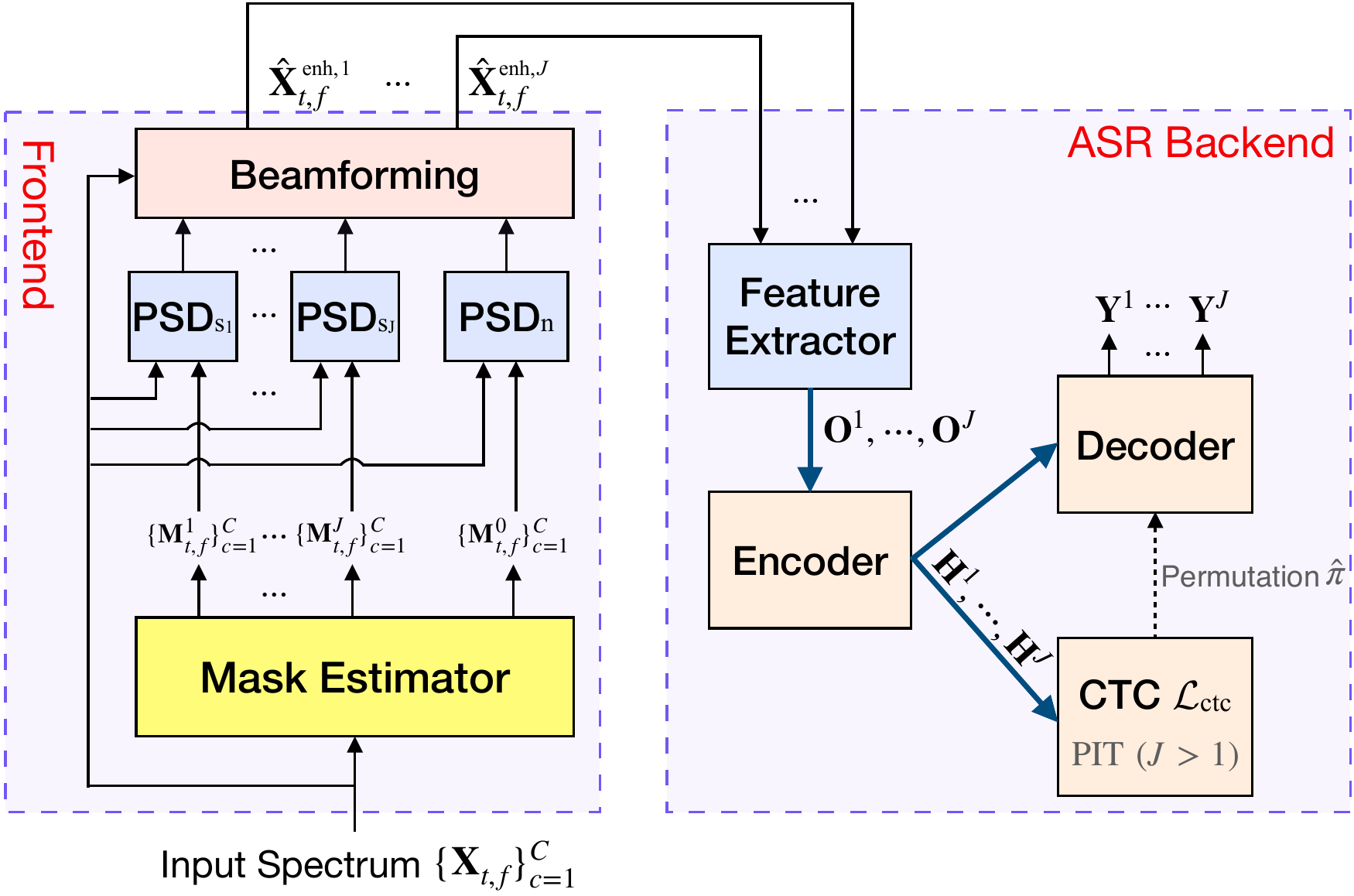}}
  \end{minipage}
\caption{End-to-End Multi-channel \gls{asr} Model.}
\label{fig:MIMO_Speech}
\end{figure}

This section reviews the end-to-end multi-channel speech recognition system for both single-speaker $(J = 1)$ \cite{Subramanian-arxiv-2019} and multi-speaker $(J > 1)$ \cite{MIMO-Chang2019, End-Chang2020} conditions, as shown in Fig.~\ref{fig:MIMO_Speech}. Without loss of generality, we consider the input speech as a mixture of $J \ (J \ge 1)$ different speakers. For simplicity, we consider the noise as the $0$-th source ($j=0$) in the input signal. 

The model is composed mainly of two modules, namely the frontend for speech separation and the backend for ASR. The frontend is a mask-based multi-source neural beamformer. First, the masking network estimates the masks $\mathbf{M}_c^j$ for every source $j \in \{0, 1, \dots, J\}$ on each channel $c \in \{1, \dots, C\}$ from the input spectrum  $\mathbf{X}_c = \left(x_{t, f, c}\right)_{t, f} \in \mathbb{C}^{T \times F}$:
\begin{align}
    \resizebox{0.9\columnwidth}{!}{
    $\mathbf{M} = \left(m_{t, f, c}^{j}\right)_{t, f, c, j} = \text{MaskEstimator}(\mathbf{X}) \in \mathbb{C}^{T \times F \times C \times (J+1)} \,,$
    }
\end{align}
where $m_{t, f, c}^{j} \in[0,1]$, $T$ and $F$ represent time and frequency dimensions respectively.
Second, the multi-source neural beamformer separates the mixture into $J$ streams using the \gls{mvdr} formalization \cite{Optimal-Souden2009}. The estimated masks are used to compute the cross-channel \gls{psd} matrices $\mathbf{\Phi}^{j}$ \cite{NTT-Yoshioka2015,Neural-Heymann2016,erdogan2016improved} and then the time-invariant filter $\mathbf{g}^{j}_f$ for each speaker~$j$:
\begin{align}
    \mathbf{\Phi}^{j}_f &= \frac{1}{\sum_{t=1}^T m^j_{t,f}} \sum_{t=1}^T m^{j}_{t,f} \mathbf{x}_{t,f} \mathbf{x}^{H}_{t,f} \; \in \mathbb{C}^{C \times C} \,, \\
    \mathbf{g}^j_f &= \frac{(\sum _{i \neq j} \mathbf{\Phi}^{i}_f)^{-1} \mathbf{\Phi}^{j}_f}{\text{Trace}((\sum _{i \neq j} \mathbf{\Phi}^{i} _f)^{-1} \mathbf{\Phi}^{j}_f)} \mathbf{u} \; \in \mathbb{C}^{C} \label{eq:mvdr-filter} \,,
\end{align}
where $\mathbf{x}_{t,f} = \{x_{t,f,c}\}_{c=1}^C$, $m^{j}_{t,f} = \frac{1}{C} \sum_{c=1}^C m^{j}_{t,f,c}$, $(\cdot)^H$ represents the conjugate transpose, and $\mathbf{u} \in \mathbb{R}^{C}$ is a vector denoting the reference microphone estimated by an attention mechanism \cite{Multichannel-Ochiai2017}.
Finally, the separated speech $\hat{\mathbf{X}}^{\mathrm{enh},j}$ of each speaker $j$ is derived by applying the filters $\mathbf{g}^{j}$ to the input speech $\mathbf{X}$, from which the log Mel-filterbank feature with global mean and variance normalization ($\text{GMVN-LMF}(\cdot)$) is further extracted:
\begin{align}
    \hat{x}^{\mathrm{enh}, j}_{t,f} &= (\mathbf{g}^{j}_f)^{H} \mathbf{x}_{t,f} \; \in \mathbb{C} \,, \\
    \mathbf{O}^{j} &= \text{GMVN-LMF}(|\hat{\mathbf{X}}^{\mathrm{enh},j}|) \,,
\end{align}
where $\hat{\mathbf{X}}^{\mathrm{enh},j} \!\in\! \mathbb{C}^{T \times F}$, and $\mathbf{O}^j$ is the extracted feature for \gls{asr}.

The backend is a joint \gls{ctc}/attention-based encoder-decoder model \cite{Joint-Kim2017} for single-channel speech recognition. First, the encoder transforms the feature $\mathbf{O}^{j}=\{\mathbf{o}^{j}_1,\dots,\mathbf{o}^{j}_T\}$ of each speaker $j$ into a high-level representation $\mathbf{H}^{j}=\{\mathbf{h}^{j}_1,\dots,\mathbf{h}^{j}_L\} \ (L \leq T)$ with subsampling. 
Then, the representation is processed by the attention-based decoder to generate the output token sequences $\mathbf{Y}^{j}=\{y^{j}_1,\dots,y^{j}_N\}$. The \gls{asr} process is formulated as follows:
\begin{align}
    \mathbf{H}^{j} &= \text{Encoder}(\mathbf{O}^j) \,, \\[-1ex]
    \mathbf{y}^{j}_n &\sim \text{Attention-Decoder} (\mathbf{H}^{j}, \mathbf{y}^{j}_{n-1}) \,,
\end{align}
where $\mathbf{y}_n^j$ is the posterior probability vector for the $n$-th token.
Note that in the multi-speaker case, i.e.~$J > 1$, in order to solve the label ambiguity problem, the \gls{pit} technique \cite{Permutation-Yu2017, Multitalker-Kolbaek2017,End2End-Seki2018,Improving-Zhang2020} is further applied in the \gls{ctc} module to determine the order of the label sequences.
The whole model is optimized with only the ASR loss $\mathcal{L}$ combining the attention and CTC losses with the determined label sequence order.

\vspace{-.8em}
\section{End-to-End ASR with Unified Frontend}
\label{sec:MIMO-REVERB-SPEECH}
\vspace{-.3em}
In this section, we introduce the proposed multi-channel speech recognition architecture for coping with the reverberant speech. 
First, we describe the mask-based WPE model for multi-channel dereverberation. 
Then, we show a cascade integration method which incorporates the mask-based WPE model followed by the model introduced in last Section, where the WPE filter coefficients are estimated for each speaker. 
Furthermore, another frontend architecture extending the \gls{wpd} beamformer \cite{Unified-Nakatani2019} is designed, which unifies the dereverberation and beamforming modules with our new formulation.
\vspace{-.6em}
\subsection{Mask-based WPE model}
\label{ssec:dnn_wpe}
The mask-based \gls{wpe} algorithm \cite{Neural-Kinoshita2017} is introduced in this subsection. First, the input spectrum $\mathbf{X} = (\mathbf{x}_{t,f})_{t,f}$ is fed into a neural network to estimate a time-frequency mask $\mathbf{m} = (m_{t,f,c})_{t,f,c}$, as formulated below:
\begin{align}
    \mathbf{m} &= \mathrm{MaskEstimator}(\mathbf{X}) \qquad\qquad \in \mathbb{R}^{T \times F \times C} \label{eq:mask} \,,\\
    \mathbf{x}_{t, f} &= \mathbf{x}^{\prime}_{t, f} + \mathbf{n}_{t, f} \approx \sum_{j=1}^J \mathbf{v}^j_f s^j_{t, f} + \mathbf{n}_{t, f} \quad \in \mathbb{C}^{C}\,, \label{eq:signal}
\end{align}
where $\mathbf{x}^{\prime}_{t, f}$ is the direct path and early reflection of the source signal, $s^j_{t, f}$ is the $j$-th source signal, $\mathbf{v}_f = [v_{f}^{(0)}, v_{f}^{(1)}, \cdots,$ $v_{f}^{(C-1)}]^T \in \mathbb{C}^{C}$ is the steering vector, and $\mathbf{n}_{t, f} \in \mathbb{C}^{C}$ is the noise and late reverberation of source signals.

With the estimated mask, the time-variant power $\mathbf{\lambda}_{t,f} $ of the input signal can be estimated by Eq.~(\ref{eq:psd}), and then the signal can be dereverberated via a standard WPE procedure:
\begin{align}
    \mathbf{\lambda}_{t,f} &= \frac{1}{C}\sum_{c=1}^C \frac{m_{t,f,c}}{\frac{1}{T}\sum_{\tau=1}^T m_{\tau,f,c}} \left|x_{t,f,c}\right|^2 \quad \in \mathbb{R} \label{eq:psd} \,, \\
    \hat{\mathbf{x}}^{\mathrm{wpe}}_{t,f} &= \mathrm{WPE}\left(\mathbf{x}_{t,f}, \mathbf{\lambda}_{t,f}\right) \label{eq:wpe} \,,
\end{align}

\subsection{Cascaded dereverberation and beamforming}
\label{sssec:arch1}
\begin{figure}[t]
  \begin{minipage}[b]{\linewidth}
    \centering
    \centerline{\includegraphics[width=0.8\textwidth]{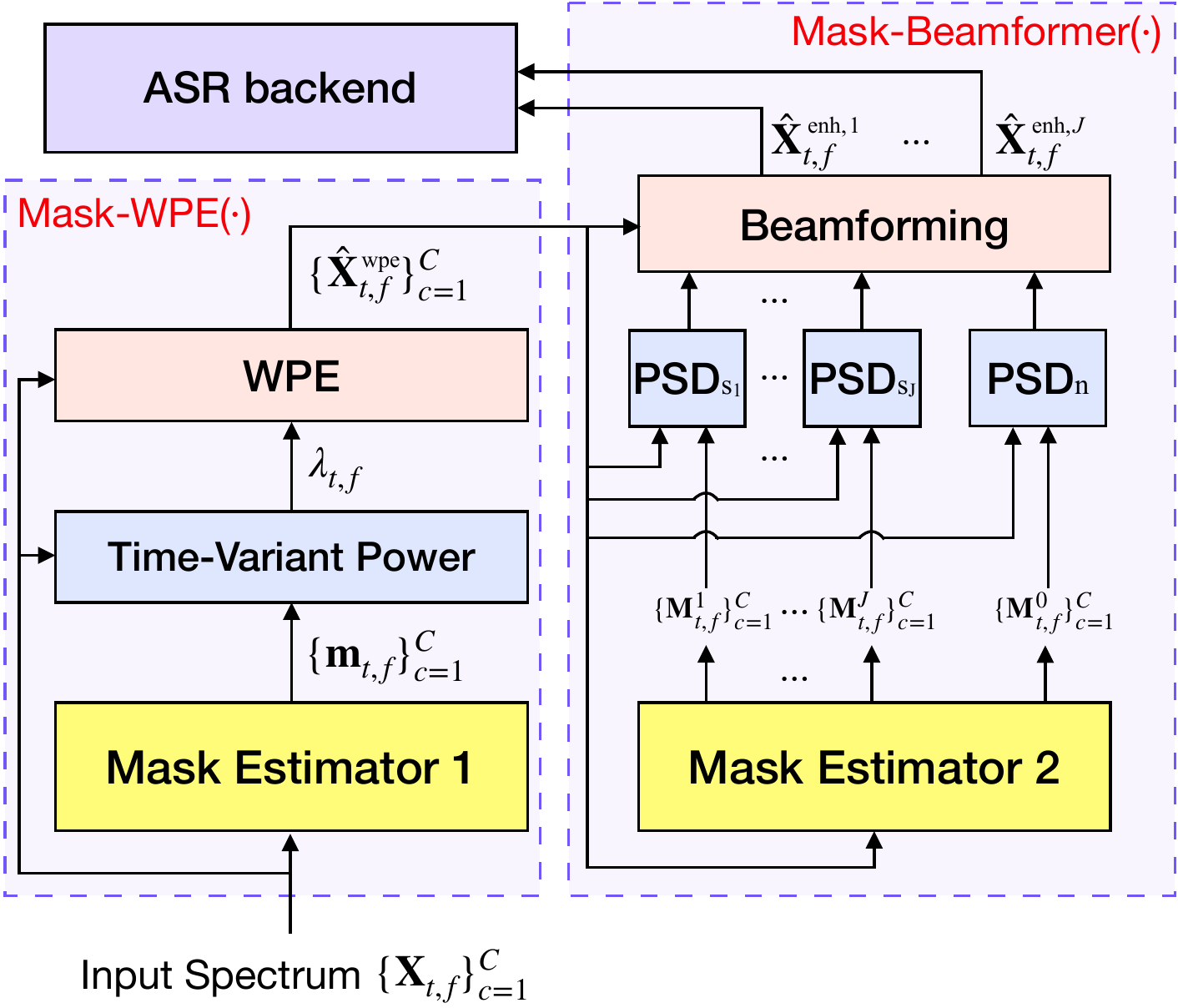}}
  \end{minipage}
\caption{Proposed end-to-end ASR arch\#1: cascaded dereverberation and beamforming frontend.}
\label{fig:MIMO_Reverb_Speech}
\end{figure}
One straightforward way to enable dereverberation in the multi-channel \gls{asr} system in Section~\ref{sec:MIMO-Speech} is the cascade integration of the mask-based WPE model and the neural beamformer like \cite{Speech-Subramanian2019}. As illustrated in Fig.~\ref{fig:MIMO_Reverb_Speech}, the multi-channel input speech mixture is first fed into the mask-based WPE model, which is composed of a mask estimator and a WPE filter. Then the dereverberated speech is processed by the beamformer introduced in Section \ref{sec:MIMO-Speech} to generate the enhanced single-channel speech of $J$ speakers for speech recognition. The frontend process can be formulated as follows:
\begin{align}
    \hat{\mathbf{X}}^{\text{enh}} = \text{Mask-Beamformer}\left(\text{Mask-WPE}\left(\mathbf{X}\right)\right) \,,
\end{align}
where $\hat{\mathbf{X}}^{\text{enh}} = \{\hat{\mathbf{X}}^{\text{enh},j}\}_{j=1}^J   \in \mathbb{C}^{T \times F \times J}$ is the set of the separated speech from all speakers, $\text{Mask-Beamformer}(\cdot)$ and $\text{Mask-WPE}(\cdot)$ denote the respective modules in Fig.~\ref{fig:MIMO_Reverb_Speech}. The ASR backend here is the same as that described in Section \ref{sec:MIMO-Speech}.

\begin{figure}[t]
  \begin{minipage}[b]{\linewidth}
    \centering
    \centerline{\includegraphics[width=0.76\textwidth]{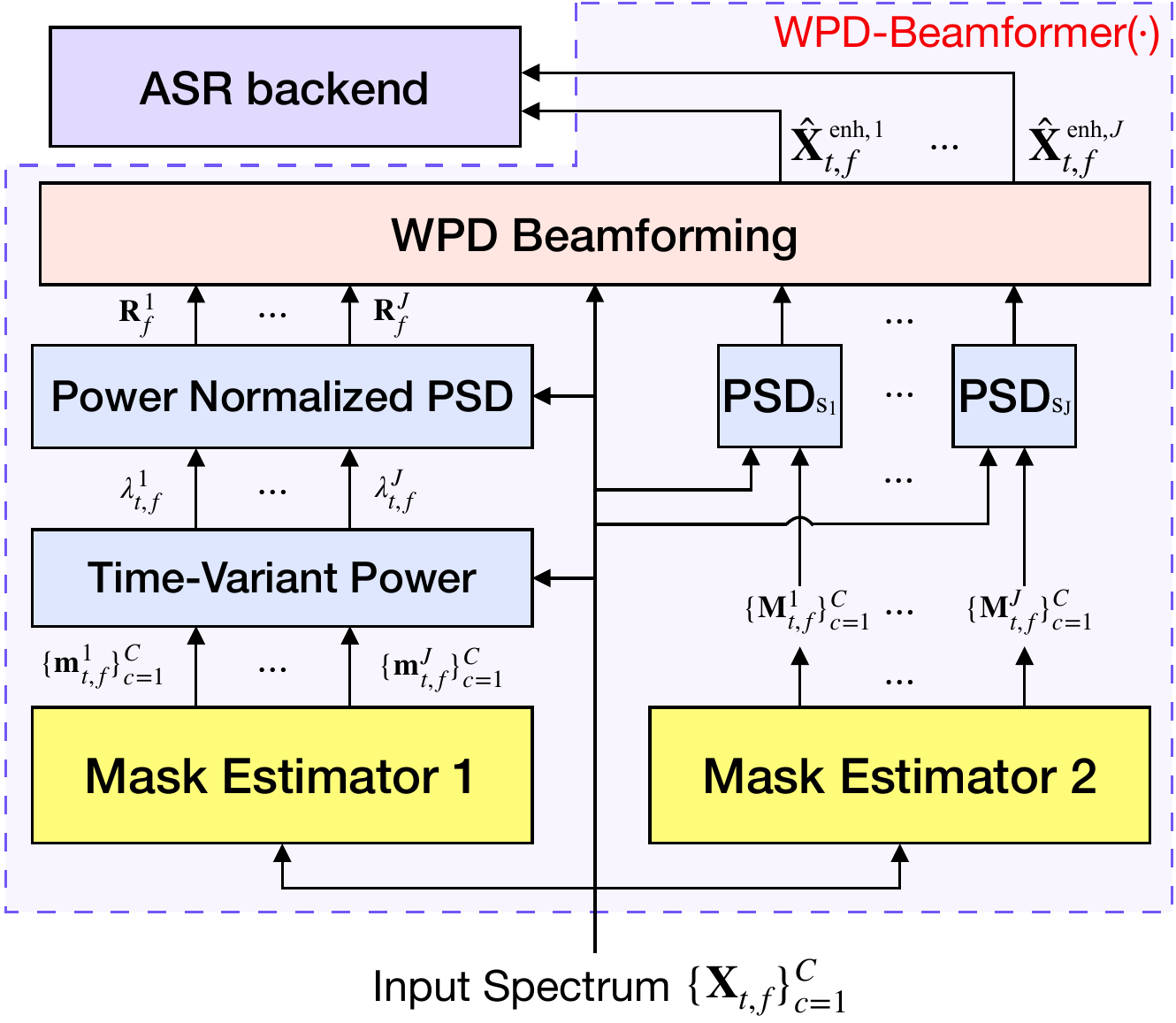}}
  \end{minipage}
\caption{Proposed end-to-end ASR arch\#2: unified dereverberation and beamforming frontend.}
\label{fig:MIMO_WPD}
\end{figure}
\subsection{Unified dereverberation and beamforming}
\label{sssec:arch2}

The original \gls{wpd} beamformer \cite{Unified-Nakatani2019} aims to eliminate the late reverberation and noise from the noisy signal, while keeping the direct signal undistorted. It combines the ideas of \gls{wpe} and \gls{mpdr} beamformer \cite{Optimum-Van2004}, and optimizes their filters at the same time, with the constrained optimization objective below:
\begin{align}
    \bar{\mathbf{w}}=\argmin _{\bar{\mathbf{w}}} \sum_{t} \frac{\left|\bar{\mathbf{w}}_{f}^{H} \bar{\mathbf{x}}_{t, f}\right|^{2}}{\lambda_{t, f}} \quad \text { s.t. } \quad \mathbf{w}_{0, f}^{H} \mathbf{v}_{f}=v_{f}^{(\text{ref})}
\end{align}
where $\bar{\mathbf{w}} = [\mathbf{w}_{0,f}^T, \mathbf{w}_{D,f}^T, \mathbf{w}_{D+1,f}^T, \cdots, \mathbf{w}_{K+D-1,f}^T]^T \in \mathbb{C}^{C(K+1)}$ is the \gls{wpd} filter coefficient, $\bar{\mathbf{x}}_{t,f} = [\mathbf{x}_{t,f}^T, \mathbf{x}_{t-D,f}^T,$ $\mathbf{x}_{t-D-1,f}^T, \cdots, \mathbf{x}_{t-K-D+1,f}^T]^T \in \mathbb{C}^{C(K+1)}$ is the concatenation of input signals of current and previous frames, $D$ is the delay parameter, $K$ is the number of filter taps, $v_{f}^{(\text{ref})}$ is the value of the steering vector at the reference channel, and $\lambda_{t,f}$ is the power of the desired signal as in Eq.~(\ref{eq:psd}).

By solving the above constrained optimization problem, we can calculate the \gls{wpd} filter $\bar{\mathbf{w}}_{f}$ and the enhanced signal $\hat{\mathbf{X}}^{\text{enh}}$ by the following formulas:
\begin{align}
    \bar{\mathbf{w}}_{f} &= \frac{\mathbf{R}_{f}^{-1} \bar{\mathbf{v}}_{f}}{\bar{\mathbf{v}}_{f}^{H} \mathbf{R}_{f}^{-1} \bar{\mathbf{v}}_{f}} \left(v_f^{(\text{ref})}\right)^* & \in \mathbb{C}^{C(K+1)} \label{eq:wpd} \,,\\
    \mathbf{R}_f &= \sum_{t=D}^{T} \frac{\bar{\mathbf{x}}_{t-D,f} \bar{\mathbf{x}}_{t-D,f}^{H}}{\lambda_{t,f}} & \in \mathbb{C}^{C(K+1) \times C(K+1)} \label{eq:corr_mat} \,, \\
    \hat{\mathbf{X}}_{t, f}^{\text{enh}} &= \bar{\mathbf{w}}_{f}^H \bar{\mathbf{x}}_{t, f} & \in \mathbb{C} \,,\qquad\quad \label{eq:wpd_enh}
\end{align}
where $\mathbf{R}_f$ is the power normalized covariance matrix, $\bar{\mathbf{v}}_f = [\mathbf{v}^T_f, \mathbf{0}, \cdots, \mathbf{0}]^T \in \mathbb{C}^{C(K+1)}$, and $(\cdot)^*$ denotes complex conjugate.
While the original \gls{wpd} can perform denoising and dereverberation simultaneously with an elegant formulation, it is only designed for speech enhancement of the single-speaker input. In addition, the steering vector $\bar{\mathbf{v}}$ in Eq.~(\ref{eq:wpd}) is needed for calculating the beamformer weights, which requires the direction information of the sound source or needs to be approximated by eigenvalue decomposition of a complex matrix \cite{Simultaneous-Nakatani2019}.

Based on the above formulation, we first derive another equivalent formula that no longer requires the steering vector $\bar{\mathbf{v}}_f$, and then extend the original \gls{wpd} to the multi-speaker case. Consider the padded speech signal $\tilde{\mathbf{x}}^{\prime T}_{t, f} = [\mathbf{x}^{\prime T}_{t, f}, \mathbf{0}, \cdots, \mathbf{0}]^T \in \mathbb{C}^{C(K+1)}$, it is easy to derive from Eq.~(\ref{eq:signal}) that:
{\allowdisplaybreaks
\begin{align}
    \tilde{\mathbf{x}}^{\prime}_{t, f} &= \bar{\mathbf{v}}_f s_{t, f} \,, \label{eq:stacked_obv} \\
    \!\!\!(\mathbf{\Phi}_{\tilde{\mathbf{x}}^{\prime}\tilde{\mathbf{x}}^{\prime}})_{f} &= \sum_{t=1}^T \dfrac{m_{t,f} \tilde{\mathbf{x}}^{\prime}_{t, f} \tilde{\mathbf{x}}_{t, f}^{\prime H}}{\sum_{\tau=1}^T m_{\tau,f}} = \bar{\mathbf{v}}_f \phi_{f} \bar{\mathbf{v}}_f^H \!=\! \bar{\mathbf{v}}_f \bar{\mathbf{v}}_f^H \phi_{f} \,, \label{eq:cov_stacked} \\
    v_f^{\text{(ref)}} &= \bar{\mathbf{v}}^T_f \bar{\mathbf{u}}  \,, \label{eq:ref_ch}
\end{align}}
where $(\mathbf{\Phi}_{\tilde{\mathbf{x}}^{\prime}\tilde{\mathbf{x}}^{\prime}})_f \in \mathbb{C}^{C(K+1) \times C(K+1)}$ is the cross-channel \gls{psd} matrix of the padded speech signal $\tilde{\mathbf{x}}^{\prime}_{t, f}$, $\bar{\mathbf{u}} = [\mathbf{u}^T, \mathbf{0},$ $\cdots, \mathbf{0}]^T \in \mathbb{R}^{C(K+1)}$ and $\mathbf{u} \in \mathbb{R}^{C}$ is the reference vector denoting the reference microphone estimated by an attention mechanism. Substitute Eq.~(\ref{eq:stacked_obv}) -- (\ref{eq:ref_ch}) into Eq.~(\ref{eq:wpd}), we can derive that:
\begin{align}
    \bar{\mathbf{w}}_f = \dfrac{\mathbf{R}^{-1}_f (\mathbf{\Phi}_{\tilde{\mathbf{x}}^{\prime}\tilde{\mathbf{x}}^{\prime}})_{f}}{\operatorname{Trace}\left[\mathbf{R}^{-1}_{f} (\mathbf{\Phi}_{\tilde{\mathbf{x}}^{\prime}\tilde{\mathbf{x}}^{\prime}})_{f}\right]} \bar{\mathbf{u}}  \label{eq:new_wpd} \,.
\end{align}
This new formula is equivalent to Eq.~(\ref{eq:wpd}), but no longer requires the steering vector for calculating the filter weights.

Furthermore, we can easily extend \gls{wpd} to the multi-speaker case. For each speaker $j$, the corresponding covariance matrix $\mathbf{R}_f^{j}$ can be derived from Eq.~(\ref{eq:corr_mat}) and (\ref{eq:psd}), where the estimated mask $\mathbf{m}^j$ for speaker $j$ is used for both dereverberation and beamforming. Then the \gls{wpd} beamformer for speaker $j$ can be calculated from Eq.~(\ref{eq:new_wpd}). Finally, the separated speech of each speaker can be derived from Eq.~(\ref{eq:wpd_enh}), using the corresponding \gls{wpd} beamformer.
Note that the masks for calculating $\mathbf{R}_f^{j}$ and $\mathbf{\Phi}_{\tilde{\mathbf{x}}^{\prime}\tilde{\mathbf{x}}^{\prime}}^{j}$ for each speaker $j$ can be either shared using a single mask estimator or estimated by two separate mask estimators.
The \gls{wpd} based architecture is illustrated in Fig.~\ref{fig:MIMO_WPD}.
\vspace{-1.3em}
\section{Experiments}
\label{sec:exp}

To make our experimental results comparable to previous results of MIMO-Speech \cite{MIMO-Chang2019,End-Chang2020}, we evaluated the proposed methods on the same spatialized wsj1-2mix dataset as in \cite{MIMO-Chang2019,End-Chang2020}, which consists of two sub-datasets: anechoic and reverberant.
The reverberation time (RT$_{60}$) of the reverberant data ranges from 200 ms to 600 ms. In each sub-dataset, the duration of the spatialized speech for training, development and evaluation is 98.5 hr, 1.3 hr and 0.8 hr respectively. We also adopt the multi-condition training in \cite{Does-Ochiai2017, MIMO-Chang2019}, i.e. include the WSJ train\_si284 in training to improve the performance.
We also test our methods for single-speaker speech recognition on the REVERB dataset \cite{kinoshita2013reverb}, which uses 2-channel simulated reverberant data for training and 8-channel real data for evaluation.

For feature extraction, the \gls{stft} is performed with a 16-kHz sampling rate and a 25-ms Hann window with a 10-ms stride, and the spectral feature's dimension is $F=257$. After the frontend processing, 80-dimensional log Mel-filterbank features are extracted from the enhanced spectrum of each separated speech, where the global mean and variance normalization is applied using the statistics from the single-speaker WSJ1 training set. The number of channels for training in our experiments is $C=2$. But it can be extended to an arbitrary number of channels as described in \cite{Multichannel-Ochiai2017}.

\vspace{-.8em}
\subsection{Experimental Setup}
\label{ssec:setup}

All our proposed end-to-end multi-channel speech recognition models are implemented based on the ESPnet framework \cite{ESPnet-Watanabe2018}.\footnote{Our experimental setup is available at \url{https://github.com/Emrys365/espnet/blob/wsj1_mix_spatialized/egs/wsj1_mix_spatialized/asr1/}.}
The AdaDelta optimizer with $\rho=0.95$ and $\epsilon=10^{-8}$ is used for training. The data in both anechoic and reverberant conditions are used for training.

In the mask-based \gls{wpe} module, the mask estimation network is a 3-layer bidirectional long-short term memory with projection (BLSTMP) network with 300 cells in each direction. The number of iterations for performing mask-based WPE is set to 1. The prediction delay $D$ and the number of taps $K$ is set to 3 and 5 respectively. The mask estimators in both the \gls{mvdr} beamformer and the \gls{wpd} beamformer are 3-layer BLSTMP networks with 512 cells.
Note that in all our experiments except Section \ref{ssec:exp3}, we used a shared mask estimator instead of two separate ones in Fig.~\ref{fig:MIMO_WPD}.
In the \gls{asr} module, following the configurations in \cite{End-Chang2020}, we use the CNN-Transformer based encoder, which consists of 2 CNN blocks and 12 Transformer layers, and the 6-layer Transformer-based decoder. The self-attention in all Transformer layers has the same configuration as in \cite{End-Chang2020}, i.e.~4 heads and 256 dimensions. As for decoding, a word-level language model \cite{End-Hori2018} trained on the official text data included in the WSJ corpus is used. 
The interpolation factor between CTC and attention losses is set to 0.2.

For experiments on REVERB, the network configuration and experimental conditions are the same as \cite{Subramanian-arxiv-2019}. The reference microphone is fixed as the second channel. Both dereverberation and denoising subnetworks are trained to predict two dimensional time-frequency masks.

\subsection{Evaluation of the proposed architectures for multi-speaker speech recognition}
\label{ssec:exp1}

\begin{table}[t]
    \caption{Performance (WER [\%]) of the proposed arch1\,/\,arch2 models with different numbers of filter taps ($K$) and microphones ($C$) on the spatialized reverberant wsj1-2mix eval set.}
    \centering
    \begin{tabular}{c|ccc}
        \toprule
        \diagbox[height=1.6\line]{$K$}{$C$} & 2 & 4 & 6 \\
        \hline
        1 & 28.87\,/\,27.44 & 17.95\,/\,16.67 & 14.92\,/\,\textbf{13.97} \\
        3 & 27.62\,/\,26.42 & 16.65\,/\,15.95 & 14.63\,/\,14.23 \\
        5 & \textbf{21.88}\,/\,25.54 & 15.93\,/\,\textbf{15.72} & 15.46\,/\,16.81 \\
        7 & 26.62\,/\,25.68 & 16.09\,/\,16.32 & 18.67\,/\,22.40 \\
        10 & 26.64\,/\,25.81 & 18.67\,/\,19.55 & 27.79\,/\,36.28 \\
        \bottomrule
    \end{tabular}%
    \label{tab:exp2}
\end{table}
\begin{table}[t]
    \caption{Performance evaluation on the spatialized reverberant wsj1-2mix corpus.}
    \centering
    \resizebox{\columnwidth}{!}{%
    \begin{tabular}{lcc}
        \toprule
        \formattitle{Model} & \formattitle{dev WER (\%)} & \formattitle{eval WER (\%)} \\
        \hline
        baseline (RNN backend) \cite{MIMO-Chang2019} & 34.98 & 29.99 \\
        $\ \ +$ Nara-WPE \cite{End-Chang2020} & 24.45 & 17.67 \\
        baseline (Transformer backend) \cite{End-Chang2020} & 32.95 & 28.01 \\
        $\ \ +$ Nara-WPE \cite{End-Chang2020} & \textbf{19.17} & \textbf{15.24} \\
        \hline
        proposed arch 1 & 19.37 & 14.63 \\
        proposed arch 2\footnotemark & \textbf{18.34} & \textbf{13.97} \\
    \bottomrule
    \end{tabular}%
    }
    \label{tab:exp1.1}
\end{table}
\footnotetext{The arch 2 model in Table~\ref{tab:exp1.1} was trained on the basis of a pretrained MIMO-Speech model, since the direct training of arch 2 currently does not work well due to numerical instability issues.}

Since our proposed architectures can be tested flexibly with different numbers of microphones $C$ and filter taps $K$, even if the model is trained with fixed $C$ and $K$, we first evaluate the performance of the proposed two architectures with different numbers of filter taps and microphones for inference on the reverberant wsj1-2mix dataset. The results are presented in Table~\ref{tab:exp2}.
We can observe that the performance can be significantly improved when more microphones are available. The number of filter taps closer to the training setup ($K=5$) usually leads to better results, but with more microphones, using fewer filter taps may also provide enough information for dereverberation and increases the stability of the operations in Eq.~(\ref{eq:new_wpd}).
The best performance for arch 1 and arch 2 is achieved with $C=6, K=3$ and $C=6, K=1$ respectively.

Then we compare the performances of our proposed models with the baseline models. In Table \ref{tab:exp1.1}, the baselines are the MIMO-Speech models with RNN backend (row 1) and Transformer backend (row 3) from our previous study \cite{MIMO-Chang2019, End-Chang2020}.
Since these baseline models do not contain a dereverberation module, we also introduce two enhanced baselines (row 2 \& 4), i.e.~MIMO-Speech with iterative Nara-WPE\footnote{\url{https://github.com/fgnt/nara_wpe}} preprocessing.
We ran Nara-WPE with 10 filter taps for 5 iterations to preprocess both training and evaluation data for the baseline models.
By comparing the four baselines and our proposed two architectures with best results taken from Table~\ref{tab:exp2}, we can observe that both proposed models combining neural dereverberation and beamforming in the end-to-end structure achieve comparable results to the best baseline ones.
Note that our models do not need an iterative process compared to the Nara-WPE preprocessing baselines.
Finally, our proposed arch 2 model based on WPD outperforms all baseline methods.

\subsection{Effectiveness of unified filtering for single source robust speech recognition}
\label{ssec:exp3}
\revised{We also evaluated our methods in the single source condition on REVERB.}
We first trained a single source multi-channel E2E ASR model, which is a variant of a cascaded architecture in Section \ref{sssec:arch1} based on the WPE and MVDR frontend \cite{Subramanian-arxiv-2019}.
Then, we replace the cascaded frontend with the proposed unified WPD frontend, and compare both frontends.
Using our WPD unified filter gives a significant improvement in performance over the cascade configuration, as shown in Table \ref{tab:result_reverb}. This shows that using our unified frontend is also effective for single source data.

These results indicate that our proposed E2E multi-channel speech recognition model is a powerful method for applications in reverberant single-speaker and multi-speaker scenarios.

\begin{table}[tb]
    \centering
    \caption{ASR performance on REVERB evaluation real dataset comparing unified and cascade filtering
    with $K=5$ \& $C=8$.}
    \label{tab:result_reverb}
    \centering
    \begin{tabular}{lcc}
        \toprule
     \formattitle{Frontend} & \formattitle{Near WER (\%)} & \formattitle{Far WER (\%)} \\
     \hline
WPE + MVDR & 10.8 & 13.6       \\
proposed arch 2 & \textbf{8.9} & \textbf{11.1}   \\\bottomrule

    \end{tabular}
\end{table}

\vspace{-.3em}
\section{Conclusion}
\label{sec:conclusion}

In this paper, we proposed an end-to-end multi-channel far-field speech recognition framework with unified dereverberation and beamforming, which is capable of performing speech dereverberation, separation and recognition simultaneously. The whole model is optimized via only the \gls{asr} criterion but can still learn relatively good dereverberation and separation skills. Two novel frontend architectures are explored, and promising performance is achieved on the spatialized wsj1-2mix corpus compared to the previous MIMO-Speech model. Experimental results on REVERB dataset also demonstrate the effectiveness of our proposed \gls{wpd} based architecture.

\vspace{-.3em}
\section{Acknowledgement}
\label{sec:ack}
This work was supported by the China NSFC project No. U1736202. Experiments have been carried out on the PI supercomputers at Shanghai Jiao Tong University.
We would like to thank the NTT Communication Laboratories for the use of their DNN-WPE module\footnote{\url{https://github.com/nttcslab-sp/dnn_wpe}} for our implementation.

\vfill
\pagebreak
\bibliographystyle{IEEEtran}
\bibliography{mybib}

\end{document}